\documentclass[12pt,preprint]{aastex}

\slugcomment{To appear in the Astrophysical Journal Letters\\
{\it submitted 21 August 2006; accepted 12 September 2006}}

\shorttitle{T\lowercase{r}ES--2} \shortauthors{O'Donovan et al.}

\newcommand{\gsc}{\mbox{\object[GSC 03549-02811]{GSC\,03549--02811}}}
\newcommand{\trs}{\mbox{\object[TrES-2]{TrES--2}}}
\newcommand{\trp}{\mbox{\object[TrES-2]{TrES--2}}}

\begin{document}

\title{T\lowercase{r}ES--2: The First Transiting Planet in the
  \boldmath{$Kepler$} Field\altaffilmark{1} }

\author{Francis~T.~O'Donovan\altaffilmark{2},
  David~Charbonneau\altaffilmark{3,4},
  Georgi~Mandushev\altaffilmark{5}, Edward~W.~Dunham\altaffilmark{5},
  David~W.~Latham\altaffilmark{3}, Guillermo~Torres\altaffilmark{3},
  Alessandro~Sozzetti\altaffilmark{3,6},
  Timothy~M.~Brown\altaffilmark{7,8}, John~T.~Trauger\altaffilmark{9},
  Juan~A.~Belmonte\altaffilmark{10}, Markus~Rabus\altaffilmark{10},
  Jos\'{e}~M.~Almenara\altaffilmark{10},
  Roi~Alonso\altaffilmark{11,10}, Hans~J.~Deeg\altaffilmark{10},
  Gilbert~A.~Esquerdo\altaffilmark{3,12},
  Emilio~E.~Falco\altaffilmark{3},
  Lynne~A.~Hillenbrand\altaffilmark{2},
  Anna~Roussanova\altaffilmark{13},
  Robert~P.~Stefanik\altaffilmark{3}, Joshua~N.~Winn\altaffilmark{13}}

\altaffiltext{1}{Some of the data presented herein were obtained at
  the W.M. Keck Observatory, which is operated as a scientific
  partnership among Caltech, the University of California, and NASA.
  The Observatory was made possible by the generous financial support
  of the W.M. Keck Foundation.}

\altaffiltext{2}{California Institute of Technology, 1200 E.
  California Blvd., Pasadena, CA 91125; ftod, lah@astro.caltech.edu}

\altaffiltext{3}{Harvard--Smithsonian Center for Astrophysics, 60
  Garden St., Cambridge, MA 02138; dcharbonneau, 
  dlatham, gtorres, asozzetti, efalco, rstefanik@cfa.harvard.edu}

\altaffiltext{4}{Alfred P. Sloan Research Fellow}

\altaffiltext{5}{Lowell Observatory, 1400 West Mars Hill Rd.,
  Flagstaff, AZ 86001; gmand, ted.dunham@lowell.edu}

\altaffiltext{6}{INAF--Osservatorio Astronomico di Torino, 10025 Pino
  Torinese, Italy}

\altaffiltext{7}{Las Cumbres Observatory Global Telescope, 6720
  Cortona Dr. Ste. 102, Goleta, CA 93117; tbrown@lcogt.net}

\altaffiltext{8}{High Altitude Observatory/National Center for
  Atmospheric Research, 3080 Center Green, Boulder, CO 80301}

\altaffiltext{9}{Jet Propulsion Laboratory, 4800 Oak Grove Dr., MS
  183-900, Pasadena, CA 91109; john.t.trauger@jpl.nasa.gov}

\altaffiltext{10}{Instituto de Astrof{\'\i}sica de Canarias, 38200 La
  Laguna, Tenerife, Spain; jba, mrabus, jmav, hdeeg@iac.es}

\altaffiltext{11}{Laboratoire d'Astrophysique de Marseille, Traverse
  du Siphon, 13376 Marseille 12, France; roi.alonso@oamp.fr}

\altaffiltext{12}{Planetary Science Institute, 1700 East Fort Lowell
  Rd. Ste. 106, Tucson, AZ 85719; esquerdo@psi.edu}

\altaffiltext{13}{Department of Physics, Massachusetts Institute of
  Technology, 77 Massachusetts Ave., Cambridge, MA 02139; 
  roussano, jwinn@mit.edu}

\begin{abstract}

We announce the discovery of the second transiting hot Jupiter
discovered by the Trans-atlantic Exoplanet Survey. The planet, which
we dub \trp, orbits the nearby star \gsc\ every 2.47063 days.  From
high-resolution spectra, we determine that the star has $T_{\rm
  eff}=5960\pm100\,\mathrm{K}$ and $\log{g}=4.4\pm0.2$, implying a
spectral type of G0V and a mass of $1.08^{+0.11}_{-0.05}\,M_{\sun}$.
High-precision radial-velocity measurements confirm a sinusoidal
variation with the period and phase predicted by the photometry, and
rule out the presence of line-bisector variations that
would indicate that the spectroscopic orbit is spurious.  We estimate
a planetary mass of $1.28^{+0.09}_{-0.04}\,M_{\rm Jup}$.  We model
$B$, $r$, $R$, and $I$ photometric timeseries of the 1.4\%-deep
transits and find a planetary radius of $1.24^{+0.09}_{-0.06}\,R_{\rm
  Jup}$.  This planet lies within the field of view of the NASA
\textit{Kepler} mission, ensuring that hundreds of upcoming transits
will be monitored with exquisite precision and permitting a host of
unprecedented investigations.
\end{abstract}



\section{Introduction}
\label{sec:intro}

Observations of the ten known transiting hot Jupiters have provided
precise planetary radii and masses, and tested formation and structure
models for extrasolar planets
\citep[see][]{Laughlin_Wolf_Vanmunster:apj:2005a,
  Charbonneau_Brown_Burrows:PPV:2006a}. More detailed studies of the
nearby planets have probed their atmospheres and led to the direct
detection of their thermal emission
\citep[e.g.][]{Charbonneau_Brown_Noyes:apj:2002a,
  Charbonneau_Allen_Megeath:apj:2005a,
  Deming_Brown_Charbonneau:apj:2005a,
  Deming_Seager_Richardson:nat:2005a}.

Three of these planets were known from radial-velocity surveys of the
solar neighborhood, and were subsequently observed to transit.  The
remaining seven were discovered from photometric observations. The
radial-velocity confirmation of transiting planet candidates involves
extensive use of large-aperture telescopes. With the goal of
maximizing the yield of transiting planets around bright stars and
minimizing the time required of large observatories, several teams are
undertaking wide-field photometric surveys using small telescopes
\citep[for a review, see][]{Charbonneau_Brown_Burrows:PPV:2006a}. Our
collaboration is conducting the
\anchor{http://www.astro.caltech.edu/~ftod/tres/}{Trans-atlantic
  Exoplanet Survey}%
\footnote[14]{\url{http://www.astro.caltech.edu/$\sim$ftod/tres/}}%
\ (TrES): TrES--1 was the first nearby transiting planet to be
discovered photometrically \citep{Alonso_Brown_Torres:apjl:2004a}.

Such photometric surveys yield numerous transit candidates, of which
the majority are astrophysical false positives that are discarded by
follow-up photometric and spectroscopic observations
\citep[e.g.,][]{ODonovan_Charbonneau_Alonso:ApJ-submitted:2006a}. 
However,
eliminating a blend, wherein a bright star forms a chance
superposition or a hierarchical triple with a faint eclipsing binary,
can require a careful analysis
\citep[][]{Torres_Konacki_Sasselov:apj:2004b,
  Mandushev_Torres_Latham:apj:2005a,
  ODonovan_Charbonneau_Torres:apj:2006a}.

We present here the discovery of the planet \trp, 
and describe the process by which we
confirmed its planetary nature and deduced its bulk properties.

\section{Observations and Analysis}
\label{sec:observations}

Transits of the parent star \trs\ were first observed by
\anchor{http://www.astro.caltech.edu/~ftod/tres/sleuth.html}{Sleuth}
(Palomar Observatory, California) and PSST (Lowell Observatory,
Arizona; \citealt{Dunham_Mandushev_Taylor:pasp:2004a}), part of the
\anchor{http://www.astro.caltech.edu/~ftod/tres/}{TrES} network of
10--cm telescopes. The third telescope
\anchor{http://www.hao.ucar.edu/public/research/stare/stare.html}{STARE}
\citep{Alonso_Deeg_Brown:an:2004a} in Tenerife, Spain, did not observe
because it was undergoing an upgrade at the time. The two telescopes
monitored a $5.7\degr \times 5.7\degr$ field of view (FOV) centered on 
the star \mbox{16\,Lyr} from UT 2005 June 16 to September 3. 
The analysis of TrES images has been described in detail in
\cite{Dunham_Mandushev_Taylor:pasp:2004a} and
\cite{ODonovan_Charbonneau_Alonso:ApJ-submitted:2006a,ODonovan_Charbonneau_Torres:apj:2006a}. In summary,
we analyzed the Sleuth and PSST images separately. After calibration,
we obtained a list of the field stars in each image, and determined
their equatorial coordinates. We applied our image spatial
interpolation and subtraction routines based in part upon
\cite{Alard:aas:2000a} to obtain the differential magnitude of each
star in each image. We decorrelated and binned the stellar light curves, 
before applying the transit-search
algorithm of \cite*{Kovacs_Zucker_Mazeh:aa:2002a} to identify stars
showing statistically-significant, periodic transit-like events.

\begin{figure}
\epsscale{0.8}
\plotone{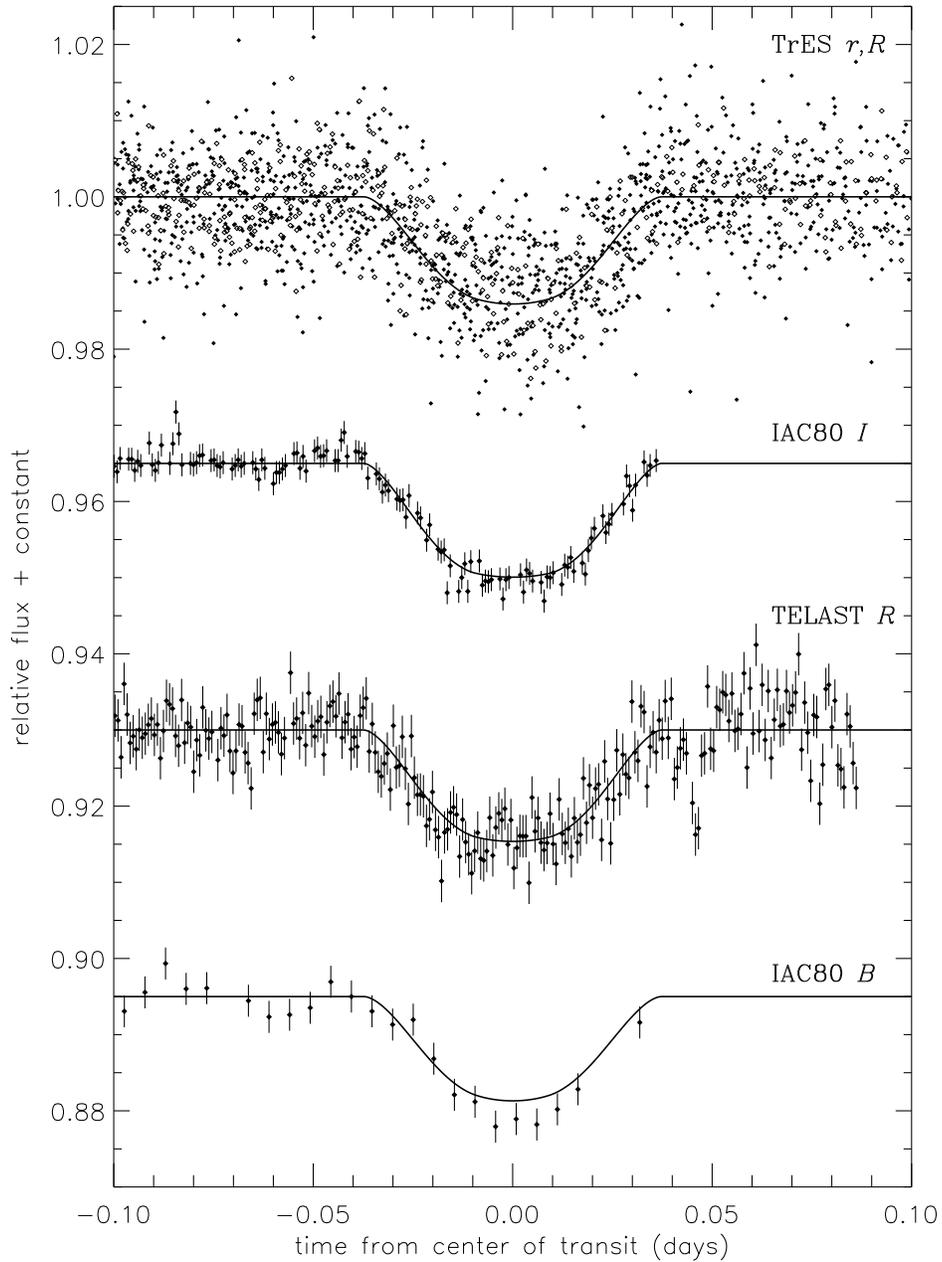}
\caption{Relative flux of the TrES--2 system as a function of time
  from the center of transit, assuming the ephemeris in
  Table~\ref{tab:tres2b}. The top light curve shows the unbinned
  discovery data, consisting of points from Sleuth $r$ ({\it solid
    diamonds}) and PSST $R$ ({\it open diamonds}). Each of the
  follow-up light curves is labeled with the telescope and filter
  employed.  We have overplotted the simultaneous best-fit solution,
  assuming the appropriate quadratic limb-darkening parameters for
  each band pass.}
\label{fig:treslc}
\end{figure}

We quickly selected \trs\ as a prime candidate. The Sleuth $r$ and
PSST $R$ photometric time series obtained near-transit and folded with
a period $P=2.47063$\,d are shown in Figure~\ref{fig:treslc}. Five
full transits and three partial transits were observed by Sleuth. PSST
observed two full transits and one partial event, events that were
also observed by Sleuth. We were therefore confident that the events
were not the result of instrumental error. The depth of 1.4\% was
consistent with the transit of a Jupiter-sized object across a
solar-type star, and the duration of only 1.5\,h implied a
near-grazing eclipse.

We searched for the counterpart of \trs\ in publicly-available
catalogs, and identified the star as \gsc. 
The 2MASS $J-K=0.386$ is consistent with a Sun-like star. 
The UCAC2
proper motion ($5.60\,\mathrm{mas\,yr^{-1}}$) is also consistent with,
but slightly less than, the expectation for a nearby dwarf. We examined
the DSS images and found no nearby bright companions within the 
$30\arcsec$ radius of the Sleuth photometric aperture.
In order to obtain absolute photometry and colors of \trs, we observed it
 in Johnson $UBV$ and Cousins $R$ 
on the nights of UT 2006 August 29 and  
30 with the 105--cm Hall telescope at Lowell Observatory.
We calibrated the data using six standard fields \citep{Landolt:aj:1992a}, 
and the results are given in Table~\ref{tab:tres2}.

We observed
\trs\ using the CfA Digital Speedometers \citep{Latham:ASP:1992a} on
UT 2005 October 18, 20, 23, November 13 and 2006 June 13.
These spectra are centered on 5187\,\AA, and cover 45\,\AA\ with
a resolving power of $\lambda / \Delta \lambda \approx 35,\!000$. By
cross-correlating these spectra with synthetic spectra created by
J.~Morse using Kurucz model stellar atmospheres (J.~Morse \&
R.~L.~Kurucz, 2004, private communication), we computed the radial
velocity (RV) at each epoch. Within the measurement error 
($\sim0.5\,\mathrm{km\,s^{-1}}$), the RVs are
constant with a mean velocity of $-0.56\,\mathrm{km\,s^{-1}}$ and a 
scatter of $0.55\,\mathrm{km\,s^{-1}}$.
This limits the mass of the companion to be less than 8\,$M_{\rm Jup}$. 
From a similar cross-correlation analysis, we
estimate (assuming a solar metallicity) the stellar effective
temperature $T_{\rm eff}$, surface gravity $\log{g}$, and the
projected rotational velocity $v \sin{i}$ (Table~\ref{tab:tres2}).
These estimates are consistent with the G0V spectral type implied by
the photometry.

We gathered rapid-cadence, high-precision photometric observations in
$I$ and $B$ on UT 2006 August 10 with the CCD camera at the IAC80, an
80--cm telescope of the Observatorio del Teide, Tenerife, Spain. The
CCD camera has a FOV of $10\arcmin\times10\arcmin$, corresponding to
$0\farcs33$~pixel$^{-1}$. After calibrating the images, we carried out
aperture photometry with VAPHOT \citep{Deeg_Doyle:2001a} on the target
and several reference stars of similar brightness in the FOV.  We
constructed an ensemble average of the calibrators, divided the target
by the resulting time series, and renormalized the resulting light
curve by the median of its value prior to the transit event.
Simultaneous $R$ observations were gathered with the TELAST 0.35--m
telescope, also located at the Teide Observatory, and were analyzed in
a similar fashion.  This telescope is able to follow the target to
larger airmass permitting greater time coverage, but the resulting
light curve showed a residual trend that was likely due to an
imperfect extinction correction. To correct for this, we fit a cubic
polynomial in time to the out-of-transit data, extended the fit across
the complete dataset, and divided the data by this function. For each
dataset, we estimated the measurement errors from the rms variation of
the data preceding first contact.  The light curves are presented in
Figure~\ref{fig:treslc}.

In order to confirm the planetary nature of the companion and measure
its mass, we carried out RV observations using Keck/HIRES
\citep{Vogt_Allen_Bigelow:SPIE:1994a} with its $\mathrm{I}_2$
absorption cell \citep{Marcy_Butler:pasp:1992a}.  Eleven star+iodine
spectra and one template spectrum were collected UT 2006 August 2--4,
permitting good sampling of critical orbital phases. We reduced the
data using the MAKEE package written by T. Barlow. Our spectra were
gathered with a resolving power of $\lambda / \Delta \lambda \simeq
71,\!000$, and with exposure times of 15~min, permitting a typical
signal-to-noise ratio of $120\,\mathrm{pixel^{-1}}$.
Our analysis procedure to derive relative RVs incorporates the full modeling of
temporal and spatial variations of the HIRES instrumental profile (%
\citealt{Valenti_Butler_Marcy:pasp:1995a}, see also 
\citealt{Butler_Marcy_Williams:pasp:1996a,
  Korzennik_Brown_Fischer:apj:2000a,
  Cochran_Hatzes_Paulson:aj:2002a}). We model each echelle order
containing $\mathrm{I}_2$ lines independently, and then calculate the
internal uncertainties for this
star) for each observation as the RV scatter about the mean divided 
by the square root of the number of spectral orders. The RV
precision achieved by our code is described in
\cite{Alonso_Brown_Torres:apjl:2004a} and
\cite{Sozzetti_Torres_Latham:preprint:2006a,
  Sozzetti_Yong_Carney:aj:2006a}. The RV measurements are listed in
Table~\ref{tab:rvtres2}.

\begin{figure}
\epsscale{1.0}
\plotone{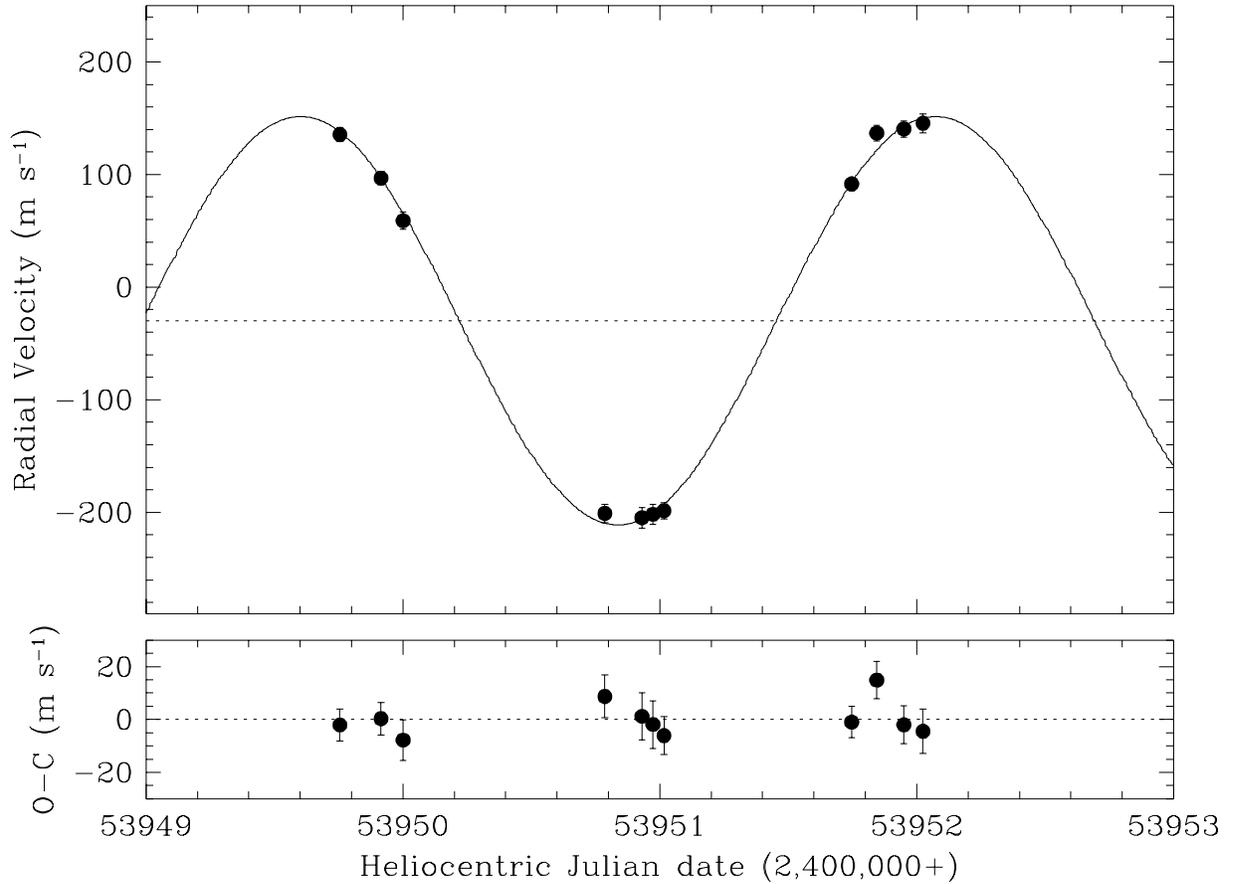}
\caption{(Top panel) Radial-velocity observations of TrES--2 obtained
  with Keck/HIRES using the $\mathrm{I}_{2}$ cell. The best-fit orbit
  (\textit{solid line}) and $\gamma$-velocity (\textit{dashed line})
  are overplotted. (Bottom panel) The residuals from the best-fit
  model to the radial-velocity data.}
\label{fig:rvtres2}
\end{figure}

The best-fit orbital solution, constrained to have zero eccentricity
(as expected from theoretical arguments for a short-period planet),
and with the $P$ and transit epoch $T_{c}$ determined from the
photometric data, yields a velocity semi-amplitude $K =
181.3\pm2.6\,\mathrm{m\,s^{-1}}$ and an instrumental
${\gamma}$-velocity of ${\gamma} = -29.8 \pm 2.2\ {\rm m\,
  s^{-1}}$. The fit has a $\chi^2_\nu = 0.89$ ($\nu=9$) and the rms of the
residuals is $6.9\,\mathrm{m\,s^{-1}}$, in excellent agreement with
the internal errors.  Figure~\ref{fig:rvtres2} shows the RV data
overplotted with the best-fit model, as well as the residuals to the
fit. The parameters of the orbital solution are listed in
Table~\ref{tab:tres2b}. We find a minimum mass for the planet of
$M_{p}\, \sin{i} = 1.206 \pm 0.016\, (\frac{M_p + M_{\star}}{M_{\sun}})^{2/3}\ M_{\rm Jup}$, where $i$ is the orbital inclination and $M_{\star}$ 
is the stellar mass.  In \S3 we estimate these two quantities to obtain
$M_{p}$.  As a further check on the consistency between the
photometric and RV datasets, we fix $P$, set $e=0$, and solve for
$T_c$ (as well as $K$ and ${\gamma}$).  We find $T_{c}=2453957.6283
\pm 0.0084$, which is consistent with, but less precisely determined
than the value predicted from the photometry (Table~\ref{tab:tres2b}).

To investigate the possibility that the RV variations are due not to a
planetary companion but rather to distortions in the spectral line
profiles arising from contamination of the spectrum by an unresolved
eclipsing binary \citep[][]{Santos_Mayor_Naef:aa:2002a,
  Torres_Konacki_Sasselov:apj:2005a}, we examined the line bisectors
carefully for signs of time-varying asymmetries. We cross-correlated
each of our Keck spectra against a synthetic spectrum matching the
measured properties of the star. Line bisectors were then computed
from the cross-correlation function averaged over spectral orders not
affected by the iodine lines, which is representative of the average
spectral line profile.  Bisector spans were calculated as the velocity
difference between points selected near the top and bottom of the
bisectors \cite[][]{Torres_Konacki_Sasselov:apj:2005a}. If the
velocity variations were the result of a stellar blend, we would
expect the bisector spans to vary in phase with the photometric period
with an amplitude similar to that seen in the RVs
\citep{Queloz_Henry_Sivan:aa:2001a, Mandushev_Torres_Latham:apj:2005a}.
 Instead, we did not detect any variation exceeding the measurement uncertainties.

As an additional check we carried out detailed modeling of the TrES
photometry following \cite{Torres_Konacki_Sasselov:apj:2004b} to test
the hypothesis that the light curve is the result of blending of the
main G0 star with an unseen eclipsing binary. The properties of the
three stars (parameterized in terms of their mass) were taken from
model isochrones subject to the $T_{\rm eff}$ and $\log{g}$
constraints on the main star. An 
excellent fit to the TrES $r$--band light curve was obtained for a
triple system composed of a G--dwarf primary blended with an
eclipsing binary with individual components of spectral type M0 and
M4--M5.  In this model, the flux ratio between the G--dwarf primary
and the brightest (M0) component of the blended binary is less than
2\%, which would be undetectable in our spectra.
However, the color difference between the G0 and M0 stars is
such that we would expect the $B$ data to present an eclipse depth
half of that in the TrES bandpass, in contrast to what is observed
(Figure~\ref{fig:treslc}). (Although we note in \S\ref{sec:discuss}
that a modest color-dependent extinction error may be present in the $B$ data, 
it is both the opposite sign and of too small an amplitude to permit the 
blend decribed here.)  More generally, any blend scenario is strongly 
disfavored by the observed RV orbit and corresponding lack of bisector 
variability.

We conclude from these tests that a blend scenario is strongly
inconsistent with the data, and therefore that the star is indeed
orbited by a Jovian planet.

\section{Estimates of Planet Parameters and
  Conclusions} \label{sec:discuss}

In order to determine $M_{\star}$ and its uncertainties, we compared
our estimates of $T_{\rm eff}$ and $\log{g}$ with evolutionary models
from \cite{Yi_Demarque_Kim:apjs:2001a}, assuming solar
metallicity. For each isochrone, we identified the range of
$M_{\star}$ for which the $T_{\rm eff}$ and $\log{g}$ lay within our
1\,$\sigma$ errors.  We took the best-fit model as our estimate of
$M_{\star}$, and the span of permitted models (over all ages greater
than 500~Myr) to be our uncertainty.  We then used the resulting
value, $M_{\star} = 1.08^{+0.11}_{-0.05}\, M_{\Sun}$, and the
spectroscopic orbit (\S2) to estimate $M_{p} = 1.28^{+0.09}_{-0.04}\,
M_{\rm Jup}$.  We also evaluated the stellar radius $R_{\star}$ in a
similar fashion, and found results that were consistent with, but less
tightly constrained than that from the light-curve modeling (below). 
The uncertainty contributed by that in the evolutionary models
is less than 0.02 solar masses. Based on the absolute visual magnitude
 ($M_{V}$=4.5) predicted 
by the best-fit model, we estimate the distance to be approximately 
230\,pc. We estimate the reddening in the direction of \trs\ to be 
$E(B-V) \sim 0.05$ and the extinction to be $\sim 0.15$\,mag from 
comparison of its observed colors with the intrinsic 
colors predicted by the model.  

To estimate $R_{\star}$, $i$, and the planetary radius $R_{p}$, we
simultaneously fit our light curves using the analytical transit
curves of \cite{Mandel_Agol:apjl:2002a} and the color-dependent
quadratic limb-darkening parameters from \cite{Claret:aa:2000a}, which
were matched to the spectroscopically-estimated properties of the
star.  We identified the best-fit solution by fixing the value of
$M_{\star}$ at its best estimate, $1.08\ M_{\Sun}$, and minimizing the
${\chi}^2$ to all the photometry.  We note that the available time
series are well-described by the model, with the exception of the
IAC80 $B$, for which the in-transit data fall below the model.  We
speculate that those data, which were gathered at high airmass, may
have been imperfectly corrected for extinction, which is a larger
effect at $B$ than the other band passes.  The best-fit solution
obtains ${\chi}_{\nu}^2=1.15$ ($\nu=2065$), and 
its values for $\{R_{\star}, R_{p},i\}$ are listed in Tables~\ref{tab:tres2} 
and \ref{tab:tres2b}. The
uncertainties in these quantities are dominated by our uncertainty in
$M_{\star}$. To derive 1\,$\sigma$ errors for each of $\{R_{\star},
R_{p}, i\}$, we change the value of that parameter and fix it at a new
value, and then allow the other two parameters to float, as well as
allow for a value of $M_{\star}$ within our uncertainty. (The uncertainties in 
$P$ and $T_{c}$ are sufficiently small so as not to contribute significantly 
to the errors in $R_{\star}$, $R_{p}$, and $i$.) We repeat
this procedure until the best-fit solution produces an increase in the
${\chi}^2$ corresponding to a 1\,$\sigma$ change.  Our estimate of the
planetary radius, $R_{p} = 1.24^{+0.09}_{-0.06}\ R_{\rm Jup}$, implies
a mean density of $0.83^{+0.12}_{-0.09} \ {\rm g\, cm^{-3}}$,
indistinguishable from that of TrES-1 \citep[using the values
  from][]{Sozzetti_Yong_Torres:apjl:2004a}, despite the fact that
\trs\ is nearly twice as massive. We also note that the impact
parameter, $b = a\, \cos{i} / R_{\star} = 0.84\pm0.02$, is the largest of any
known transiting exoplanet.

We intend to improve our estimates of the planetary and stellar
parameters by undertaking a more detailed analysis of the stellar
spectrum as we did for TrES--1
\citep{Sozzetti_Yong_Torres:apjl:2004a}, and by gathering very
high-precision $z$-band photometry
\citep[e.g.][]{Holman_Winn_Latham:preprint:2006a}.  Such data will
permit us to look for transit timing variations indicative of
additional planets in the \trs\ system
\citep{Agol_Steffen_Sari:mnras:2005a, Holman_Murray:Science:2005a,
  Steffen_Agol:mnras:2005a}. \trs\ lies within the FOV of the NASA
\textit{Kepler} mission. During the four year mission,
\textit{Kepler} will observe nearly 600 transits of \trs. 
The precision with which \textit{Kepler} will observe these transits will
enable an extremely sensitive search for additional planets in the \trs\
system through their dynamical perturbations.  Moreover, the large impact
parameter means that very subtle changes in its value could be detected.
Such variations are predicted \citep{Miralda-Escude:apj:2002a} to occur as a result
of either additional planets, or the stellar quadrupole moment. \textit{Kepler}
may also detect the reflected light from \trs\ \citep{Jenkins_Doyle:apj:2003a}
and hence determine the long-sought geometric albedo and phase function of
a hot Jupiter.  The large impact parameter also makes \trs\ particularly
favorable for determining the angle between the stellar spin-axis and the
orbital axis via the Rossiter--McLaughlin effect 
\citep{Gaudi_Winn:preprint:2006a}. 
\cite{Williams_Charbonneau_Cooper:preprint:2006a} discuss
the use of {\it Spitzer} IRAC observations spanning the time of
secondary eclipse to resolve the surfaces of extrasolar planets.  The
large impact parameter of the \trs\ orbit is ideal for this
application, since it grants access to both longitudinal and
latitudinal flux variations across the dayside hemisphere of the
planet.

\acknowledgments
We sincerely thank  R.~Brucato, M.~Doyle, K.~Dunscombe, R.~Ellis, 
B.~Gordon, J.~Henning, L.~Kroll, S.~Kunsman, J.~Mueller, H.~Petrie, 
A.~Pickles, N.~Scoville, M.~Sweet, R.~Thicksten, G.~van~Idsinga, 
R.~Wetzel, and D.~Zieber for their assistance with the fabrication, 
operation, and maintenance of the Sleuth instrument.
We thank the referee, S.~Gaudi, for his detailed comments that helped 
improve the paper. 
We are indebted to S.~Fern\'{a}ndez~Acosta who accommodated the
unscheduled transit observation at the IAC80, which is
operated by the IAC in its Observatorio del Teide. 
The authors wish to recognize and acknowledge the very significant 
cultural role and reverence that the summit of Mauna Kea has always 
had within the indigenous Hawaiian community. We are most fortunate 
to have the opportunity to conduct observations from this mountain. 
This material is based upon work supported by NASA under grants 
NNG05GJ29G, NNG05GI57G, NNH05AB88I, and NNG04LG89G, 
issued through the Origins of Solar Systems Program. We acknowledge 
support from the NASA \textit{Kepler} mission.

\clearpage

\begin{deluxetable}{lcc}
\tablewidth{0pt}
\tablecaption{Parent Star \label{tab:tres2}}
\tablehead{ \colhead{Parameter} & \colhead{Value}  &  \colhead{Reference} }
\startdata
R.A. \phm{00000000.} (J2000)  &  \phm{000.}$19^{\rm h} 07^{\rm m} 14\fs03$ &  \\
Decl. \phm{00000000} (J2000)  &  $+49\arcdeg 18\arcmin 59\farcs3$ &  \\
GSC & \mbox{\object[GSC 03549-02811]{03549--02811}} & \\
$V$ \phm{00000000000.} (mag) & $11.411\pm0.005$ & a \\
$B-V$ \phm{$0000000.$} (mag) &  \phn$0.619\pm0.009$ & a \\
$U-B$  \phm{$0000000.$} (mag) &  \phn$0.112\pm0.012$ & a\\
$V-R_{\rm C}$  \phm{$000000.$} (mag) &  \phn$0.361\pm0.008$ & a\\
$J$  \phm{000000000000}  (mag) &  $10.232 \pm 0.020$ & b \\
$J-H$  \phm{0000000.} (mag) & \phn$0.312 \pm 0.033$ & b \\
$J-K_{s}$  \phm{0000000} (mag) & \phn$0.386 \pm 0.030$ & b \\
$[\mu_{\alpha},\mu_{\delta}]$ \phm{00000.} ($\mathrm{mas\ yr^{-1}}$) &  $[4.45,-3.40]$ & c \\
Spectral Type  &  G0V & a \\
$M_{\star}$ \phm{00000000000} ($M_{\sun}$)  &  $1.08^{+0.11}_{-0.05}$  & a \\
$R_{\star}$ \phm{00000000000} ($R_{\sun}$) &  $1.00^{+0.06}_{-0.04}$  & a \\
$T_{\rm eff}$ \phm{00000000000} (K) & $5960 \pm 100$ & a \\
$\log{g}$ \phm{000000000.} (dex) & $4.4 \pm 0.2$ & a \\
$v\sin{i}$ \phm{0000000.} (${\rm km\, s^{-1}}$) & $2.0 \pm 1.5$ & a
\enddata
\tablenotetext{a}{This work.}
\tablenotetext{b}{From the 2MASS Catalog.}
\tablenotetext{c}{From the UCAC2 Bright Star Supplement.}
\end{deluxetable}

\clearpage
\begin{deluxetable}{lc}
\tablewidth{0pt}
\tablecaption{TrES--2 Planet \label{tab:tres2b}}
\tablehead{ \colhead{Parameter} & \colhead{Value} }
\startdata
$P$ \phm{000.} (d) &  \phm{000000} $2.47063\pm0.00001$ \\
$T_{c}$ \phm{.} (HJD)  & $2453957.6358\pm0.0010$ \\
$a$ \phm{00.} (AU) &  \phm{000000} $0.0367^{+0.0012}_{-0.0005}$ \\
$i$ \phm{00000} ($\degr$)  &  \phm{000000} $83.90\pm0.22$  \\
$K$ \phm{.}($\mathrm{m\,s^{-1}}$) &  \phm{0000} $181.3\pm2.6$ \\
$M_{p}$  ($M_{\rm Jup}$) &  \phm{00000} $1.28^{+0.09}_{-0.04}$ \\
$R_{p}$ \phm{.}($R_{\rm Jup})$\tablenotemark{a} &  \phm{00000} $1.24^{+0.09}_{-0.06}$ \\
\enddata
\tablenotetext{a}{$R_{\rm Jup} = 71,\!492$~km, the equatorial radius
  of Jupiter at 1~bar.}
\end{deluxetable}

\clearpage
\begin{deluxetable}{lcc}
\tablewidth{0pt}
\tablecaption{Relative radial-velocity measurements of TrES--2 \label{tab:rvtres2}}
\tablehead{ \colhead{Observation Epoch} & \colhead{Radial Velocity}  &  \colhead{$\sigma_{\rm RV}$} \\
 \colhead{HJD - $2,\!400,\!000$} & \colhead{m s$^{-1}$}  &  \colhead{m s$^{-1}$}}
\startdata
53949.76054  &  \phm{$-$}135.5   & 6.1 \\
53949.91993  &   \phm{$-$0}96.8   & 6.1 \\
53950.00216  &   \phm{$-$0}58.9   & 7.7 \\
53950.79018  & $-$201.0   & 8.1 \\
53950.93491  & $-$204.8   & 9.0 \\
53950.98051  & $-$201.7   & 9.0 \\
53951.02136  & $-$198.5   & 7.2 \\
53951.75032  &   \phm{$-$0}91.7   & 6.0 \\
53951.84863  &  \phm{$-$}136.7   & 7.0 \\
53951.95209  &  \phm{$-$}140.5   & 7.1 \\
53952.02736  &  \phm{$-$}145.6   & 8.4 \\
\enddata
\end{deluxetable}

\clearpage

\clearpage

\end{document}